\documentclass[12pt]{iopart}
\usepackage[T1]{fontenc}
\usepackage[latin9]{inputenc}
\usepackage{amssymb}
\usepackage{graphicx}
\usepackage{esint}
\usepackage[english]{babel}
\usepackage[final]{pdfpages}
\usepackage{notes2bib}
\usepackage{soul}

\makeatletter
\@ifundefined{textcolor}{}
{%
 \definecolor{BLACK}{gray}{0}
 \definecolor{WHITE}{gray}{1}
 \definecolor{RED}{rgb}{1,0,0}
 \definecolor{GREEN}{rgb}{0,1,0}
 \definecolor{BLUE}{rgb}{0,0,1}
 \definecolor{CYAN}{cmyk}{1,0,0,0}
 \definecolor{MAGENTA}{cmyk}{0,1,0,0}
 \definecolor{YELLOW}{cmyk}{0,0,1,0}
}

\@ifundefined{definecolor}{\usepackage{color}}{}

\makeatother

\global\long\def\avg#1{\langle#1\rangle}

\begin{document}

\title{Tunable Thermal Switching via DNA-Based Nano Devices}

\author{Chih-Chun Chien$^1$, Kirill A. Velizhanin$^1$, Yonatan Dubi$^{2,3}$, Michael Zwolak$^4$}

\address{$^1$Theoretical Division, Los Alamos National Laboratory, Los Alamos, NM 87545 \\
$^2$Landa Laboratories, 3 Pekeris St., Rehovot 76702, Israel \\
$^3$Department of Chemistry and the Ilse Katz Center for Nano-Science, Ben-Gurion University, Beer-sheva 84105, Israel \\
$^4$Department of Physics, Oregon State University, Corvallis, OR 97331
}

\eads{\mailto{chihchun@lanl.gov}, \mailto{mpzwolak@physics.oregonstate.edu}}


\date{\today}

\begin{abstract}
DNA has a well-defined structural transition -- the denaturation of
its double-stranded form into two single strands -- that strongly
affects its thermal transport properties. We show that, according
to a widely implemented model for DNA denaturation, one can engineer
DNA ``heattronic'' devices that have a rapidly increasing thermal conductance over
a narrow temperature range across the denaturation transition ($\sim350\,\mathrm{K}$).
The origin of this rapid increase of conductance, or "switching", is the softening of the lattice 
and suppression of nonlinear effects as the temperature crosses the transition temperature and DNA denatures. Most
importantly, we demonstrate that DNA nanojunctions have a broad range
of thermal tunability due to varying the sequence and length, and
exploiting the underlying nonlinear behavior. We discuss the role
of disorder in the base sequence, as well as the relation to genomic
DNA. These results set the basis for developing thermal devices out
of materials with nonlinear structural dynamics, as well as understanding
the underlying mechanisms of DNA denaturation. 
\end{abstract}

\submitto{\NT}

\section{Introduction}
Thermal transport in nanoscale materials and molecules has enormous
potential in developing devices that manage heat in electronic and
other systems \cite{Dubi11-1}. For instance, thermal rectifiers \cite{Chang06-1,Schmotz11-1},
thermal transistors \cite{Saira07-1}, tunable thermal links \cite{Chang07-1}, and thermal memory \cite{Wang08-1,Xie11-1} have been experimentally demonstrated (for a recent review, see \cite{Li2012}). One can envision that many
more such devices will become feasible as methods are developed to
engineer and control nonlinear effects in materials that transport heat.

Nature has provided us with a versatile and diverse nonlinear
structure: DNA. The structural dynamics of DNA are fundamentally
interesting due to their relevance in biological processes, such as
transcription \cite{Chen10-1} and replication \cite{Alberts02-1}. Further, 
DNA is also being used in constructing functional nanoscale devices,
such as a template for electronic devices \cite{Diventra04-1} and
molecular motors \cite{Omabegho09-1}. Thus, its ability to transport
heat under different conditions is technologically important and may
allow the ``DNA template'' to be exploited not just as a scaffold
but also as a functional device in itself. In addition to theoretical
predictions \cite{Velizhanin11-1}, a recent experiment shows that
incorporation of DNA into a device can indeed give rise to nonlinear
behavior in the thermal current \cite{vanGrinsven12-1}. The experimental setup examines a change in the thermal conductance from a combined duplex DNA and fluid 
conductor to a disordered single-stranded DNA layer, the latter being thermally insulating compared to the former. 
A complete theoretical reconstruction of the experimental results would thus need to delineate the role of DNA's intrinsic
thermal conductance from the surrounding media, and examine a disordered layer of single-stranded DNA.

In this work, however, we envision instead a 
single molecule of duplex DNA bridging two thermal reservoirs in a {\em water vapor} atmosphere.
Such environment is essential for our purposes since (i) it suppresses solvent-mediated leakage heat currents
between the reservoirs, and yet (ii) the vapor atmosphere has been shown to preserve
the natural behavior of DNA (e.g., denaturation)  \cite{PeyrardPRL11}.  We demonstrate that in this setup one can tune the thermal transport
properties of DNA by taking advantage of its function as the carrier
of the genetic code via its sequence of the four bases \textendash{}
Adenine (A), Guanine (G), Cytosine (C), Thymine (T). The sequence of bases determines
both local structural properties that influence the thermal conductance
and also where nonlinear effects give way to denaturation. Together
with the length of the DNA strand, these characteristics make DNA's thermal transport
properties highly tunable. Based on this behavior, we predict that a DNA-based nano-device can act as a \emph{thermal switch}: 
the thermal conductance can rise rapidly by many orders of magnitude as
the temperature of the DNA strand is driven across the denaturation
transition. Thus, the proposed device can switch between "off" (i.e., heat-insulating) and "on" (i.e, heat-conducting) states. This is the ``heattronic'' analog of an electronic switch \cite{Diventra04-1}.
Further, we illustrate the ``engineering principles''
behind tuning thermal transport, which will be broadly applicable to nonlinear materials and help set the foundations for developing
novel thermal devices for applications in, e.g., nanoscale electronics.
\begin{figure}
\begin{center}
\includegraphics[width=8cm]{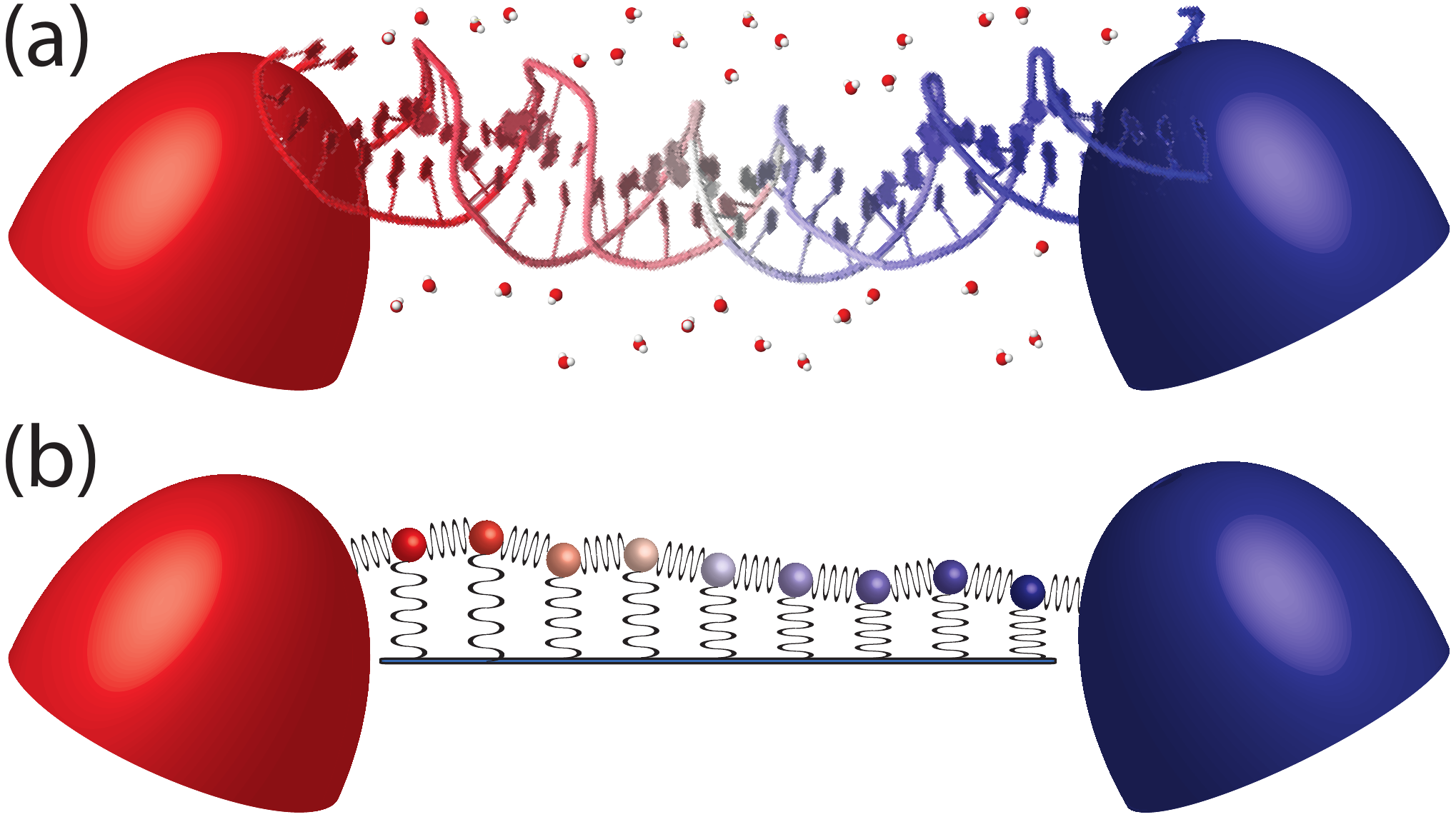}
\end{center}
\caption{\label{fig:schematic} (a) Schematic of a DNA strand (in a water vapor atmosphere) bridging two thermal reservoirs.
A hot reservoir (left, red) pumps heat into DNA, which can cause partial
denaturation, and a cold reservoir (right, blue) absorbs heat from
DNA. (b) Schematic of the PBD model for DNA dynamics. It represents DNA as a nonlinear lattice of fluctuating base 
pairs, i.e., the base to base distance in each base pair is confined in a non-linear (Morse) potential transverse to the backbone, neighboring bases 
couple only through the fluctuation of this coordinate. This does not include the helical nature of the ordered state. 
The strongly fluctuating denatured DNA is predicted to have higher
thermal conductance according to the common model for DNA denaturation.}
\end{figure}

\section{Theoretical Analysis}

Our starting point is the Peyrard-Bishop-Dauxois (PBD) model \cite{Peyrard89-1,Dauxois93-1,Dauxois93-2},
which considers double-stranded DNA as a one-dimensional lattice of
nonlinear oscillators. This model \textendash{} the common model for
the dynamics of DNA denaturation \textendash{} captures the essential
statistical features of DNA's structural transition and allows for
the direct calculation of non-equilibrium thermal transport properties
\cite{Velizhanin11-1}  
\bibnote{We note that other theoretical work has examined thermal transport
in models of DNA \cite{Terraneo02-1,Peyrard06-1,Savin11-245406}. However,
these have not examined the strong nonlinear effects during denaturation,
which is our focus here and in Ref. \cite{Velizhanin11-1}. 
}, see Figure \ref{fig:schematic}. Within the PBD model, the DNA is
described by the Hamiltonian 
\begin{equation}
H=\sum_{n}\left[\frac{m_{n}\dot{y}_{n}^{2}}{2}+V_{n}\left(y_{n}\right)+W_{n}\left(y_{n},y_{n-1}\right)\right],\label{eq:H}
\end{equation}
 where each base pair (bp) of mass $m_{n}$ is represented by stretching
of its hydrogen bonds via the coordinate $y_{n}$. The onsite and
nearest-neighbor interaction potentials, $V_{n}$ and $W_{n}$, depend
on the sequence of bases. The potentials take on the form $V_{n}(y_{n})=D^{n}(e^{-a_{n}y_{n}}-1)^{2}$ (known as the Morse potential),
which describes hydrogen bonding and effective interactions due to
the backbone/environment, and 
\begin{equation}
W_{n}(y_{n},y_{n-1})=\frac{K^{n}}{2}(1+\rho_{n}e^{-\beta_{n}(y_{n}+y_{n-1})})(y_{n}-y_{n-1})^{2},\label{eq:PBD}
\end{equation}
which describes the stacking interaction between neighboring base 
pairs. The scenario of interest is a DNA strand in a water vapor atmosphere similar to recent experiments \cite{PeyrardPRL11}, where it was shown that the PBD model still describes the transition well \bibnote{In fact, the PBD model, or a PBD-like model, may be applicable even in vacuum, as it was shown that the DNA duplex can retain some of its structural properties under extreme conditions \cite{Rueda03-1}.}. For the analytic results, we consider a uniform stacking parameter,
and then we address sequence dependent stacking using numerical simulations (see Refs. \cite{Campa98-1,Alexandrov09-1}, and the Supplemental Data for
a detailed derivation and numerical parameters). We designate the sequence of DNA by the series of bases in one of the strands (in the 3'-to-5' direction). The sequence of the other strand is unambiguously determined by requiring the DNA duplex to be 100\% complementary. Thus, only DNA double strands with no mismatches are dealt with in this work.

We will first focus on the thermal conductance ratio 
\begin{equation}
R=\frac{\kappa_{H}}{\kappa_{L}},
\end{equation}
where $\kappa_{L\left(H\right)}$ is the thermal conductance at low
(high) temperatures. This quantity was introduced in Ref. \cite{Velizhanin11-1}
as a way to characterize heat transport properties of a material near its thermally induced structural transition. Here,
we will examine its sequence dependence. When calculated using a small
temperature change, e.g., around the denaturation temperature, it
can play the role of an ``on-off'' ratio. We will see that adjusting 
the sequence and length of DNA can tune $R$ while the sequence alone allows
the transition temperature to be tuned within certain limits.

One of the main principles behind the nonlinear behavior predicted by the
PBD model is captured in the low ($L$) and high ($H$) temperature
 limits of Eq. (\ref{eq:H}): 
\begin{equation}
H_{\mu}=\sum_{n}\left[\frac{m_{n}\dot{y}_{n}^{2}}{2}+D_{\mu}^{n}y_{n}^{2}+\frac{K_{\mu}^{n}}{2}\left(y_{n}-y_{n-1}\right)^{2}\right],
\end{equation}
where $\mu=L,\: H$. We note that the PBD model is an effective model
of DNA near the denaturation transition  
\bibnote{The PBD model only includes oscillations that describe base
pair opening fluctuations. It does not include fluctuations of the
atomic constituents of the backbone. This, of course, will change
the thermal conductance. One would expect that the low temperature
thermal conductance of random sequences would be mostly due to backbone contributions, and thus the low temperature conductance should not
fall to zero but level off at some finite value. However, since the
thermal conductance is dependent on frequency, these higher frequency, localized 
modes should contribute less to the conductance for moderately sized homogeneous 
motifs. This will be the subject of a future investigation.}.
However, these limiting forms that occur at much higher/lower temperatures
give the appropriate physical description -- within the PBD model -- of DNA going from its double-stranded
to single-stranded forms. With reservoirs attach on the end sites, the thermal conductance
of an infinite strand in these limits has the form 
\begin{equation}
\kappa\equiv\frac{J}{T_{H}-T_{L}}=\int_{\omega\in W}d\omega\,\mathfrak{T}(D_{\mu}^{n},K_{\mu}^{n},m,\gamma,\omega)\label{eq:kappa},
\end{equation}
which can be calculated analytically (see Refs. \cite{Casher71-1,Dhar01-1,Velizhanin11-1} and the Supplemental
Data). Here, $J$ is the heat current, the integration is over
the frequencies that correspond to propagating modes ($W$), $\gamma$
characterizes the coupling strength to the reservoirs, and the transmission
function $\mathfrak{T}$ is determined by the structure of the lattice.

\begin{figure*}
\begin{center}
\includegraphics[width=3 in]{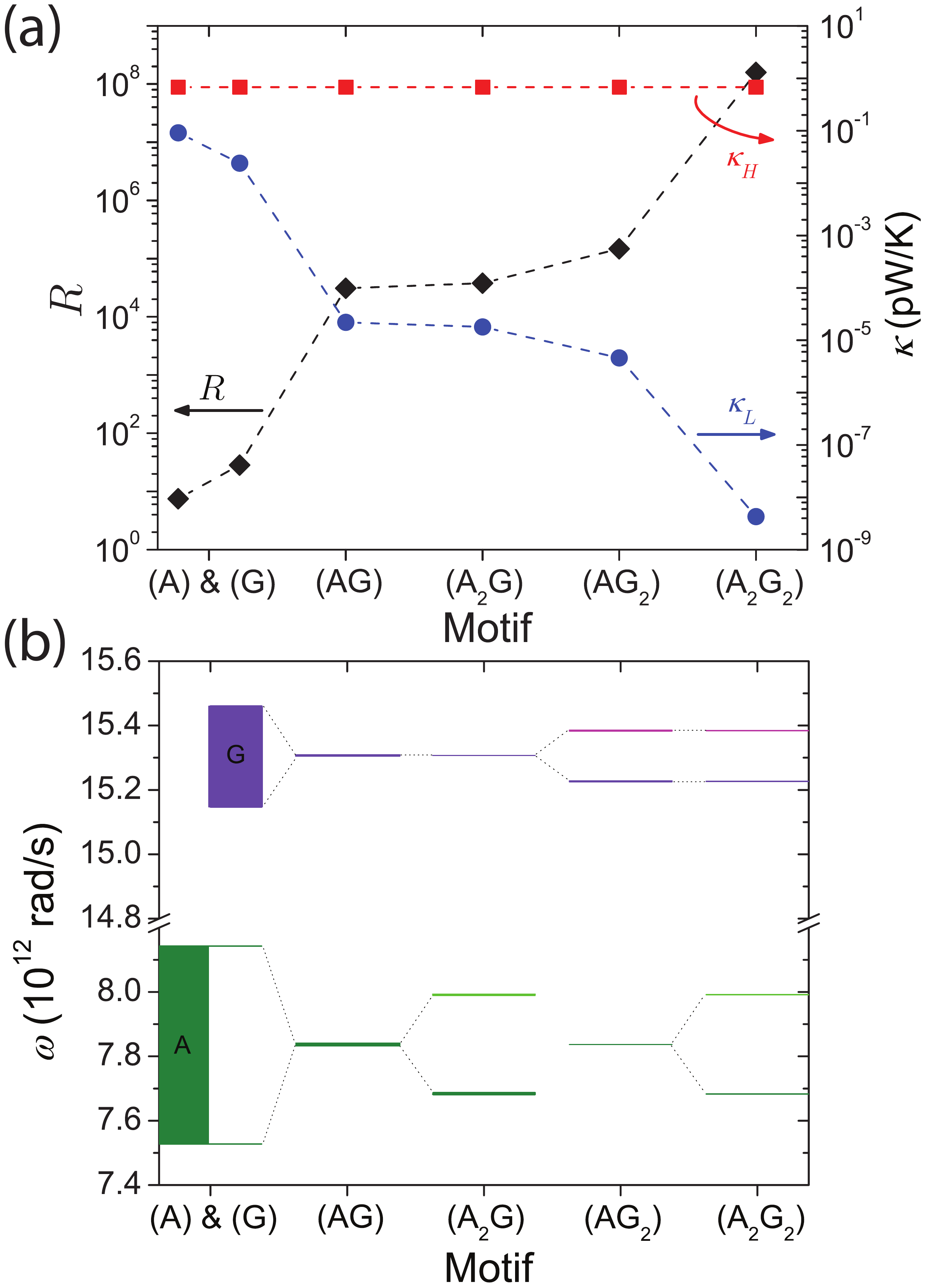}
\par\end{center}

\caption{\label{fig:Analytics} Engineering the thermal conductance of DNA. (\textbf{a})
The thermal conductance ratio and conductance for infinite 
strands with different periodic motifs computed analytically from the high/low temperature limits of
the PBD model. The low temperature conductance (blue circles), $\kappa_{L}$,
drops precipitously as the motif of the sequence is enlarged while
the high temperature conductance (red squares), $\kappa_{H}$, stays
the same. The thermal conductance ratio, $R$, introduced in Ref.
\cite{Velizhanin11-1}, is the ratio of the high to low temperature
thermal conductance and characterizes how DNA's thermal conductance
changes as it denatures. (\textbf{b}) Analytically calculated phonon bands, from the lattice parameters used to fit the PBD model to 
experimental denaturation curves, for selected DNA
sequences at low temperature. The left most bands are for a homogeneous
sequence of G (upper, purple band) and a homogeneous sequence of A
(lower, green band). As the motif is enlarged, the bands will become
more narrow and also split. In other words, periodic sequences with
increasing lengths of periodicity will decrease the bandwidth of the
phonon modes, drastically reducing DNA's ability to conduct heat at
low temperature. Although some bands have relatively small widths on the plot, all bands have finite widths.}
\end{figure*}

Figure \ref{fig:Analytics}(a) shows $\kappa_{L\left(H\right)}$ and $R$
for an infinite strand with several different motifs, i.e., the basic unit cells of the DNA lattice.
The high temperature conductance of all the sequences is identical due to the uniform stacking interaction. The low temperature
conductance, however, varies tremendously as the motif is changed and is universally much lower than its high temperature counterpart. This behavior is driven by two distinct physical mechanisms. First, going from the low temperature form to the high temperature form results in the release
of the onsite confining potential upon denaturation. This leads to softening of the phonon modes and consequently to the increase of the thermal conductance, as described analytically in Ref. \cite{Velizhanin11-1}. Second, the introduction of different motifs creates a non-uniform lattice 
(due to the different binding potentials of the AT and GC pairs). This results in a narrowing and splitting of the phonon bands, as 
shown in Fig. \ref{fig:Analytics}(b), and a subsequent reduction in the
low temperature thermal conductance. In the extreme
case of a semi-infinite poly(A) strand connected to a semi-infinite
poly(G) strand the non-uniformity would have maximal effect: The phonon bands would have no overlap (see Fig. \ref{fig:Analytics}(b))
and the strand as the whole would have zero heat conduction within this model.
Actual DNA, though, will have other contributions to heat conduction
(e.g., the backbone), which will lead to a non-zero conductance.

A natural question to ask is what is the heat conductance of genomic
or random sequences? Studying strands with periodic motifs helps understand the
behavior of random sequences. An infinite random sequence \textendash{}
and likely genomic sequences \textendash{} will look like a periodic
strand with an extremely long motif. The allowed bandwidth of propagating
modes will be narrowed by the large number of sites in the motif.
Thus, we do not expect heat to be conducted efficiently and the low-temperature
$\kappa$ should be very small for random sequences (compared to a
uniform sequence or an alternating sequence). Genomic DNA, of course,
is always finite and not completely random. We expect, however, that
the result will be similar to that of a random sequence. However,
small regions of the genome can look very different from a random
sequence, and nature may exploit sequence variation to optimize heat
(signal) transport.

\section{Numerical Results and Discussions}
The PBD model was 
developed to describe the properties of DNA around the denaturation
transition. The analytic results above addressed high and low temperature
limits (within the simplification of uniform stacking interactions).
In order to understand the extent to which the values of $R$ realized in
Fig.~\ref{fig:Analytics}(a) can be realized in a narrow temperature range
around a transition, we perform numerical simulations of the full
model including a sequence-dependent stacking interaction (parametrized
in Ref.~\cite{Alexandrov09-1}). The heat current is obtained by
keeping the temperature difference between the heat baths constant
($T_{H}-T_{L}=10~K$) and scanning the average temperature $\avg T=(T_{H}+T_{L})/2$.
We consider 90 base-pair (bp) long strands with 20 bps at each end connected
to Langevin reservoirs. The damping of the individual sites by the reservoirs is $0.5$~ps$^{-1}$. 
This is large enough to keep the very ends at the temperature of the reservoirs, 
while still allowing the sites to fluctuate at their natural frequency. 
The Supplemental Data has further details on the numerical simulations.

\begin{figure*}
\begin{center}
\includegraphics[width=3 in]{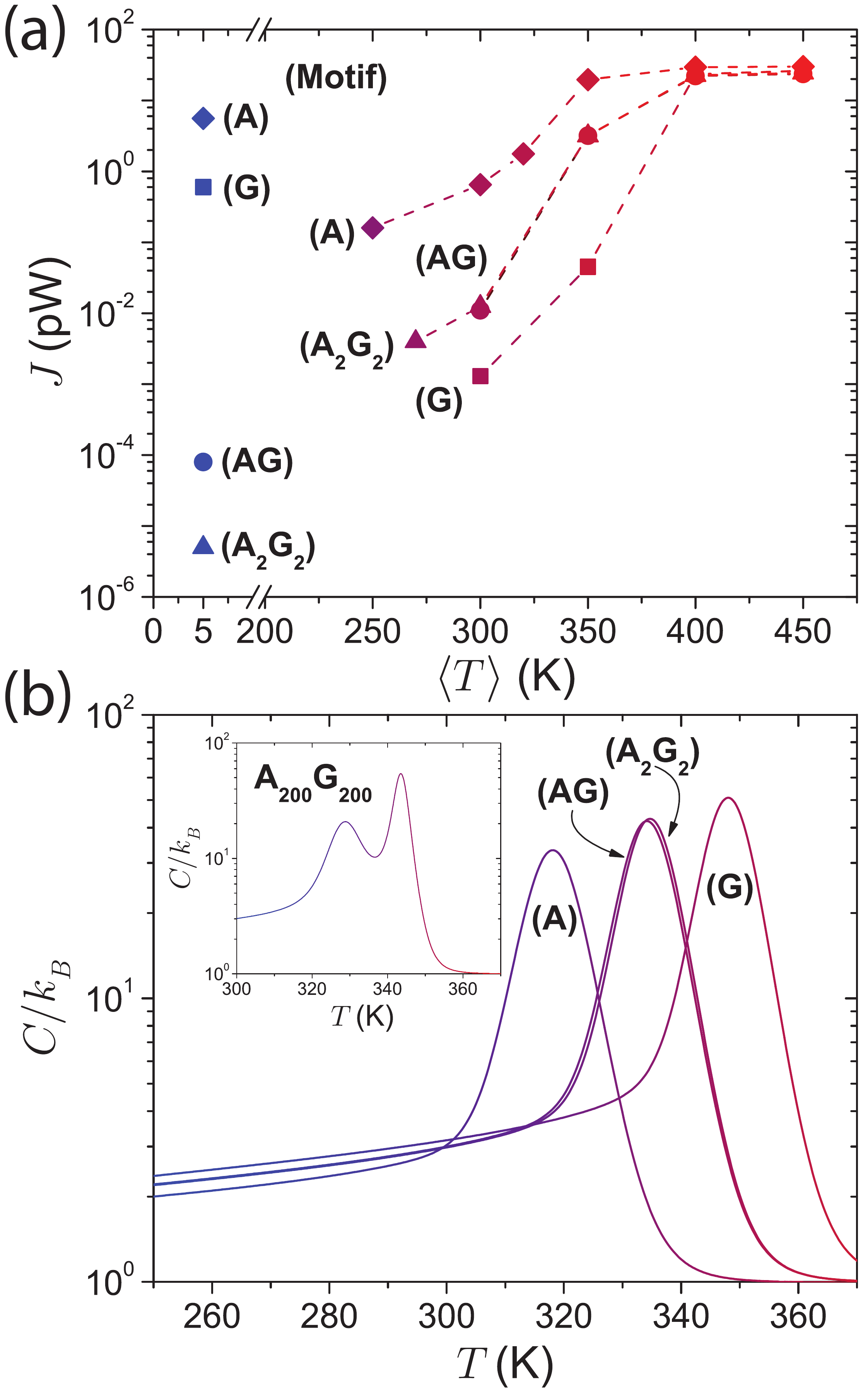}
\par\end{center}

\caption{\label{fig:OpTemp} Engineering the operation
temperature of a DNA-based thermal device using the sequence dependence
of the denaturation transition. (a) Numerically calculated heat current across select DNA
sequences with $\Delta T$=10 K and with 90 bp length, where the first
and the last 20 bp segments are connected to Langevin reservoirs.
(b) Numerically calculated heat capacity normalized per number of base pairs for a 90 base-pair (bp)
strand with periodic boundary conditions.
The inset shows the heat capacity for an infinite chain with $A_{200}G_{200}$ repeat
unit. The peak in the heat capacity marks the spot where DNA transitions into the denatured regime, and thus it sets the temperature 
around which the conductance will drastically increase.}
\end{figure*}

Fig.~\ref{fig:OpTemp}(a) shows the heat current for several sequences.
The analytical results above predicted that the heat conductance at
temperatures higher than the melting temperature is insensitive to
a particular sequence since the stacking interaction was assumed uniform.
This is not the case in the numerical simulations, where the stacking
potential assumes a more realistic sequence-dependent form.
Accordingly, the heat current exhibits a dependence (although rather weak) on sequence
in the temperature range $400-450$~K. However,
the heat conductance of DNA increases drastically when the temperature increases across 
the denaturation point. That this is indeed the denaturation transition where
the conduction of DNA strand changes rapidly (versus temperature) can be seen by the correspondence between these
curves and the peaks in the heat capacity shown in Fig.~\ref{fig:OpTemp}(b).
Around the transition, poly(AG) and poly(A$_{2}$G$_{2}$) have about
the same conductance and heat capacity, implying that the denaturation bubbles -- where
the two strands locally come apart -- are much longer than the motif and, thus, only the average sequence matters. Thus, the sequence
can be used to tune the ``operating
temperature'' of the device via its effect on the denaturation temperature. The GC base pair has a higher dissociation
energy than AT, and thus its incorporation into a strand increases
the transition temperature.

\begin{figure*}
\begin{center}
\includegraphics[width=3 in]{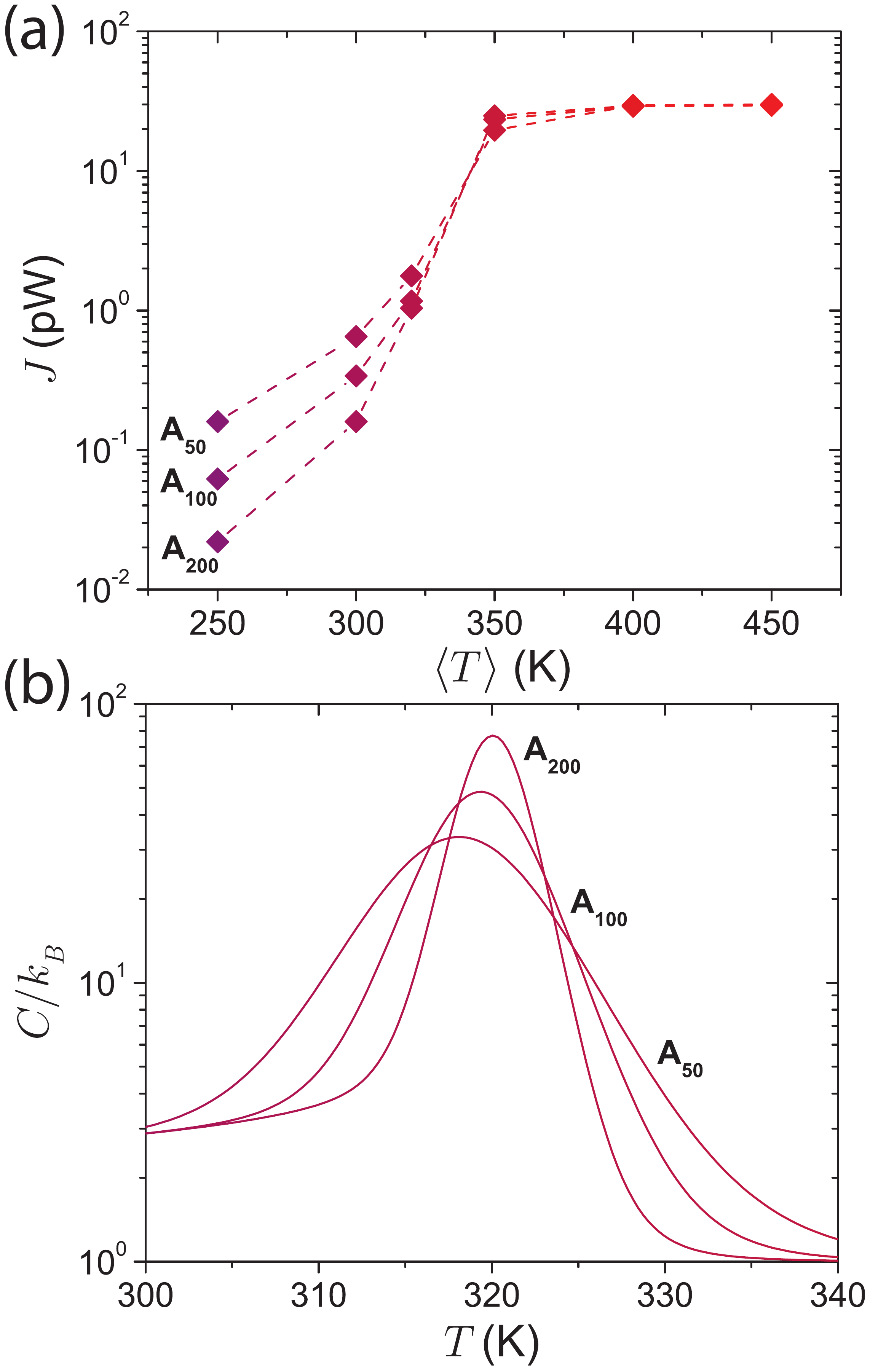}
\par\end{center}

\caption{\label{fig:Range} Engineering the range of operation
using the length of DNA. (a) Numerically calculated heat conductance of poly(A) of length
50, 100, and 200 bps (with an additional 20 bp segment connected to
a reservoir at each end). (b) Numerically calculated heat capacity around the denaturation temperature
for poly(A) of varying lengths.}
\end{figure*}

Furthermore, while the sequence can change the ``operating'' temperature and the thermal conductance ratio, 
$R$, measured by the high/low temperature limits, the on-off ratio
around the ``operating'' temperature, e.g., just below to just above the transition, is due to a more complex set of factors than just $R$. However, tuning the length of the DNA nano-junction allows one to directly tune this
important device characteristic.
Fig.~\ref{fig:Range}(a) shows
the heat conductance of poly(A) strands of various lengths.
Below the transition,
the strand is anharmonic and is expected to demonstrate finite well-defined \emph{conductivity},
i.e., the conductance is expected to be inversely proportional to
the length of the strand so long as this length is longer than the
typical bubble size. As seen in Fig.~\ref{fig:Range}(a), near (and above) the
transition, the conductance weakly depends on the length, signifying that the harmonic, high temperature Hamiltonian is being
approached. This is further supported by Fig.~\ref{fig:Range}(b), where above the transition point the heat capacity is seen to rapidly approach $C/k_B=1$ -- the harmonic limit.  

At temperatures further below the transition, the conductance
drops inversely with length of the DNA within the error of the simulations
\bibnote{We note that, right at the transition, the shorter DNA actually has
lower conductance. This is likely due to the effect of the ``clamped''
ends, which make anharmonic effects more important for the shorter
strands due to keeping the bases at the ends in the bound state.}.
This observation is in agreement with the heat capacity which shows
the transition narrowing for longer strands of DNA. This is simply
an indication that when the bubble length becomes comparable to the
strand length denaturation has effectively occurred. This finite size effect broadens the transition in temperature. 
We conjecture that using sequence effects (e.g., the suppression of
the low temperature conductance shown in Fig.~\ref{fig:Analytics}(a))
together with length will allow for even more drastic on-off jumps
in the thermal conductance. However, a more detailed study of DNA, including
backbone effects, will be required to investigate this issue.

\section{Conclusion}
To summarize, we have examined the
thermal transport properties of DNA as described by the PBD model. We predict that a DNA-based nano-device can act as a thermal switch due to its rapidly rising thermal conductance as the temperature of the DNA strand is driven across the denaturation
transition. The operating principle behind this behavior is the release of the base pairs from 
their confining potential, which both softens the lattice and suppresses nonlinear effects as the temperature 
is increased through the transition \bibnote{The latter of these is the most important for 
observing the increase before the transition: We have extensively examined the model of Joyeux and
Buyukdagli \cite{Joyeux05-1}. There are important differences between this model and the PBD model \cite{Velizhanin11-1}, but the increase in conductance before the transition is not one of them. They both predict a large increase as bubbles become larger and larger, and this strong, temperature dependent source of nonlinearity is released.}. Using analytic calculations and numerical simulations with sequence-specific parameters, we have shown that 
the operating temperature of the thermal switch can be tuned by choosing different DNA motifs and that the "on/off" ratio can be tuned 
by the DNA length. Our suggested experiments are well within current experimental reach, and recent advances in the measurement of the 
thermal conductance in various nano-junctions composed of, e.g., carbon nanotubes \cite{Chang2008}, Si nanowires \cite{Li2003,Hochbaum2008,Boukai2008}, and especially individual DNA-gold complexes \cite{Kodama2009} give potential routes to realizing the setup we propose. 

Further possibilities for engineering thermal transport may be offered
by molecular or chemical modification of the nucleotides, using much longer sequences
(e.g., see the inset of Fig.~\ref{fig:OpTemp}(b), showing that a two step
jump in conductance may be possible), and exploiting extrinsic changes
in heat conduction (e.g., due to a structural change modifying the
surrounding environment in addition to changing intrinsic properties,
as in a recent experiment \cite{vanGrinsven12-1}). This work sets the
foundation to developing thermal switches out of materials and molecules
with nonlinear structural dynamics. In addition, it will allow one to
test underlying mechanisms for structural transitions \cite{Velizhanin11-1}
and, in particular, the dynamical behavior captured within the PBD
model \cite{Velizhanin11-1,Peyrard04-1,Peyrard08-1}. We speculate
that biological systems may take advantage of such nonlinear behavior
in engineering their own control of heat flows and signaling. 

\ack{This work was performed under the NNSA of the U.S. DOE at LANL under LDRD and Contract No. DE-AC52-06NA25396, and, in part, by ONR and NAS/LPS.}

\section*{Reference}
\bibliographystyle{unsrt}



\section*{Supplementary material}

# Supplemental Material for "Tunable Thermal Switching via DNA-Based Nano Devices"

Chih-Chun Chien,[1, *] Kirill A. Velizhanin,[1, 2] Yonatan Dubi,[3, 4] and Michael Zwolak[5, †]

[1] *Theoretical Division, Los Alamos National Laboratory, Los Alamos, NM 87545*
[2] *CNLS, Los Alamos National Laboratory, Los Alamos, NM 87545*
[3] *Landa Laboratories, 3 Pekeris St., Rehovot 76702, Israel*
[4] *Department of Chemistry and the Ilse Katz Center for Nano-Science,
Ben-Gurion University, Beer-sheva 84105, Israel*
[5] *Department of Physics, Oregon State University, Corvallis, OR 97331*
(Dated: October 29, 2012)

## I. BASIC MODEL AND PARAMETERS

In the main text we discuss the thermal transport properties of a DNA nano-junction. We designate the sequence of DNA by the series of bases in one of the strands and its complementary strand is implicitly implied. Furthermore, a DNA sequence with a periodic motif is denoted by, e.g., poly(AG), which is the alternating sequence AGAGAG...

We describe the DNA in the framework of the PBD model [1–3] widely used to describe the dynamics of DNA denaturation. Essentially, the DNA is represented as a set of balls and springs with the balls representing the base-pair stretching and the springs encoding the complex interactions of bases within pairs (H-bonding) and between pairs (stacking). The PBD Hamiltonian is

$$H = \sum_n \left[ \frac{m\dot{y}_n^2}{2} + V_n(y_n) + W_n(y_n, y_{n-1}) \right]. \quad (1)$$

$y_n$ is the stretching of the base pair hydrogen bonds, and $m_n$ are the masses of the base pairs, which are assumed uniform throughout the strand with $m_n = m = 300$ a.m.u. [4, 5]. The functions $V(y_n)$ and $W(y_n, y_{n-1})$ describe the effective (intra-pair) potential felt by the hydrogen bond and the effective inter-pair stacking interaction, respectively. These potentials take the form

$$V_n(y_n) = D^n (e^{-a_n y_n} - 1)^2$$
$$W_n(y_n, y_{n-1}) = \frac{K^n}{2}(1 + \rho_n e^{-\beta_n(y_n + y_{n-1})})(y_n - y_{n-1})^2. \quad (2)$$

Recently developed parameters [4, 5] of the Morse potential for complementary base pairs are $D^{A-T} = 0.05$ eV, $a_{A-T} = 4.2$ Å$^{-1}$ for the $A-T$ base pair, and $D^{G-C} = 0.075$ eV, $a_{G-C} = 6.9$ Å$^{-1}$ for the $G-C$ base pair. The parameters of the stacking interaction between successive base pairs depends on both the pairs and their orientation. However, we first consider the uniform "average" stacking potential with parameters $K^n = K = 0.025$ eV/Å$^{-2}$, $\rho_n = \rho = 2$, and $\beta = 0.35$ Å$^{-1}$ [5]. The effect of the sequence-dependent stacking interaction is addressed in the numerical simulations using the Langevin equation.

## II. ANALYTIC DERIVATION OF THE THERMAL CONDUCTANCE

The thermal conductance of a classical harmonic lattice can be found analytically. Our starting point is to consider the limiting cases of the single-coordinate Hamiltonian (Eq. (1) in the main text) for a lattice of length $N$. The details of the high and low-temperature expansions were presented in Ref. [6]. The low- (L) and high- (H) temperature limits can be approximated by a harmonic Hamiltonian of the form

$$H_\mu = \sum_n \left[ \frac{m\dot{y}_n^2}{2} + D_\mu^n y_n^2 + \frac{K_\mu}{2}(y_n - y_{n-1})^2 \right], \quad (3)$$

where $\mu = L, H$, and the corresponding coefficients $K_H = K$, $D_H = 0$ and $K_L = K(1+\rho)$, $D_L = Da^2$.

Once a harmonic Hamiltonian is obtained, one can follow the procedure of Refs. [7, 8]. To simplify the analysis, the lattice is coupled to two heat reservoirs at the first and last sites, which gives the equations of motion

$$m\ddot{y}_n = -2(D_\mu^n + K_\mu)y_n + K_\mu(y_{n-1} + y_{n+1})$$
$$+ (\delta_{n,1} + \delta_{n,N_t}) \left[ \int_{-\infty}^t dt' A(t-t') y_n(t') + \eta_n(t) \right]. \quad (4)$$

We choose the spectrum of the dissipation to be ohmic, $A(\omega) = -i\gamma\omega$, with coupling $\gamma$, and the noise to be a white noise, $\langle \eta_{L/H}(\omega)\eta_{L/H}(\omega')\rangle = 4\pi T_{L/H}\gamma\delta(\omega+\omega')$, with $T_{L/H}$ the low and high reservoir temperatures. We consider a periodic lattice of length $N_t = N \times N_s$, where $N_s$ is the length of the motif. This form for the reservoirs will satisfy the fluctuation-dissipation theorem. The resulting equations of motion are

$$m\ddot{y}_n = -2(D_\mu^n + K_\mu)y_n + K_\mu(y_{n-1} + y_{n+1})$$
$$+ (\delta_{n,1} + \delta_{n,N_t}) [-\gamma\dot{y}_n(t) + \eta_n(t)]. \quad (5)$$

The solution for the coordinates has the form

$$y_n(t) = (1/2\pi) \int_{-\infty}^\infty d\omega \hat{Y}_{nm}^{-1}(\omega) \hat{\eta}_m(\omega) e^{i\omega t}, \quad (6)$$

where $\hat{\eta}$ is a vector of length $N_t$ with the first and last components being $\eta_{L/H}(\omega)$ and the rest being zero. It

represents the coupling of the reservoirs to the ends of the lattice. The $N_t \times N_t$ matrix $\hat{Y} = \hat{\phi} - \omega^2 \hat{M} - \hat{A}$ encodes the solution. Here $\hat{\phi}_{nm} = 2(D_\mu^n + K_\mu)\delta_{n,m} - K_\mu \delta_{n,m+1} - K_\mu \delta_{n,m-1}$, $\hat{M}_{ij} = m\delta_{i,j}$, and $\hat{A}_{11} = \hat{A}_{N_t N_t} = A(\omega)$ and $\hat{A}_{nm} = 0$ otherwise.

The heat current flowing into the lattice is $J = \langle [\int_{-\infty}^t dt' A(t-t') y_1(t')] \dot{y}_1(t) \rangle$, where the average is over the noise. Setting $\gamma = \lambda m$, the heat current becomes

$$J_\mu = \frac{\Delta T \lambda^2 m^2}{\pi} \int_{-\infty}^\infty d\omega \omega^2 \{ (\mathcal{D}_{1,N_t} - \lambda^2 \omega^2 m^2 \mathcal{D}_{2,N_t-1})^2 + \lambda^2 \omega^2 m^2 (\mathcal{D}_{1,N_t-1} + \mathcal{D}_{2,N_t})^2 \}^{-1} |C_{1,N_t}|^2, \quad (7)$$

where $\Delta T = T_H - T_R$ is the temperature difference of the reservoirs, $C_{1,N_t}$ is the cofactor of $\hat{Y}_{1,N}$, and $\mathcal{D}_{n,m}$ is the determinant of the submatrix of $(\hat{\phi} - \omega^2 \hat{M})$ from the $n$-th row (column) to the $m$-th row (column). It follows that $|C_{1,N_t}|^2 = K_\mu^{2N_t-2}$ and $\mathcal{D}_{n,m} = K_\mu^{n-m+1} \mathcal{D}_{n,m}^0$.

The elements $\begin{pmatrix} \mathcal{D}_{1,2N_t}^0 & -\mathcal{D}_{1,2N_t-1}^0 \\ \mathcal{D}_{2,2N_t}^0 & -\mathcal{D}_{2,2N_t-1}^0 \end{pmatrix} = [\mathcal{T}_s]^N$, where $\mathcal{T}_s = \mathcal{T}_1 \mathcal{T}_2 \cdots \mathcal{T}_{N_s}$ is the composite transfer matrix of one segment with length $N_s$. Since each transfer matrix $\mathcal{T}_i$ is unimodular, for an infinite lattice the allowed propagating modes correspond to $\mathcal{T}_s$ with eigenvalues $e^{\pm iq}$. We notice that for those propagating modes,

$$\mathcal{D}_{1,N_s}^0 - \mathcal{D}_{2,N_s-1}^0 = 2\cos(q). \quad (8)$$

This equation determines $\omega(q)$. In general, for a basis with $N_s$ elements, there are $N_s$ bands that satisfy Eq. (8) though the bandwidths can be substantially reduced as $N_s$ increases. Moreover, $[\mathcal{T}_s]^N = [\cos(Nq)]\mathbf{1} + [\sin(Nq)/\sin(q)][\mathcal{T}_s - (\cos(q))\mathbf{1}]$, where $\mathbf{1}$ is the $2 \times 2$ identity matrix. Then we rewrite the denominator of Eq. (7) as $|z_A|^2$, where $z_A = (1 + \lambda^2 \omega^2 m^2/K_\mu^2)\cos(Nq) + (\sin(Nq)/\sin(q))(z_c - (1 + \lambda^2 \omega^2 m^2/K_\mu^2)\cos(q))$ and $z_c = (\mathcal{D}_{1,N_s}^0 - \lambda^2 \omega^2 m^2 \mathcal{D}_{2,N_s-1}^0/K_\mu^2) - i\lambda\omega(m/K_\mu)(\mathcal{D}_{1,N_s-1}^0 + \mathcal{D}_{2,N_s}^0)$. Eq. (8) determines $\omega(q)$ so the integration of Eq. (7) can be considered as an integration over $q$ for those propagating modes. As $N \to \infty$, one uses the formula $\int_0^{2\pi} dq F(q, Nq) \to \frac{1}{2\pi} \int_0^{2\pi} dq \int_0^{2\pi} dx F(q,x)$ by treating $x = Nq$ as an independent variable in the $N \to \infty$ limit. After integrating over $x$ and $q$, one obtains the thermal current.

For a uniform lattice, $N_s = 1$ and $D_\mu^n = D_\mu$ so

$$\mathcal{T}_s = \mathcal{T}_1 = \begin{pmatrix} 2(1+D_\mu/K_\mu) - (m/K_\mu)\omega^2 & -1 \\ 1 & 0 \end{pmatrix}. \quad (9)$$

Eq. (8) gives $2\cos(q) = 2(1 + D_\mu/K_\mu) - (m/K_\mu)\omega^2$. After changing variables from $\omega$ to $q$ that satisfy this constraint, the final expression (for an infinite lattice $(N \to \infty)$) is

$$\frac{J_\mu}{\Delta T} = \frac{\gamma}{2\pi m} \int_0^{2\pi} dq \frac{\sin^2(q)}{1 + \frac{2\gamma^2}{mK_\mu}\left[1 + \frac{D_\mu}{K_\mu} - \cos(q)\right]}. \quad (10)$$

This gives for the low and high temperature thermal conductance, $\kappa_\mu = J_\mu/\Delta T$,

$$\kappa_\mu = \frac{k_B m K_\mu^2}{4\gamma^3}\left[1 + \frac{2\gamma^2}{mK_\mu} + \frac{2\gamma^2 D_\mu}{mK_\mu^2} - \mathcal{B}_\mu\right], \quad (11)$$

with

$$\mathcal{B}_\mu = \sqrt{1 + \frac{4\gamma^2}{mK_\mu} + \frac{4\gamma^2 D_\mu}{mK_\mu^2} + \frac{8\gamma^4 D_\mu}{m^2 K_\mu^3} + \frac{4\gamma^4 D_\mu^2}{m^2 K_\mu^4}}. \quad (12)$$

With these expressions one can explicitly find the thermal conductance ratio $R$. We have verified that for reservoirs contacted to a single site on each end, the thermal conductance from our numerical simulations agree with our analytic formula to within $10-15\%$. The error may be attributed to finite size effects in the numerical simulations.

We can take various limiting forms of these equations. If we define the prefactor as $\tilde{\kappa}_\mu$ and a dimensionless reservoir coupling as

$$\gamma_\mu = \frac{\gamma}{\sqrt{mK_\mu}}, \quad (13)$$

the expressions for the conductance become

$$\kappa_\mu = \tilde{\kappa}_\mu \left[1 + 2\gamma_\mu^2 + 2\gamma_\mu^2 \frac{D_\mu}{K_\mu} - \mathcal{B}_\mu\right], \quad (14)$$

with

$$\mathcal{B}_\mu = \sqrt{1 + 4\gamma_\mu^2 + 4\gamma_\mu^2 \frac{D_\mu}{K_\mu} + 8\gamma_\mu^4 \frac{D_\mu}{K_\mu} + 4\gamma_\mu^4 \left(\frac{D_\mu}{K_\mu}\right)^2}. \quad (15)$$

The appropriate limiting forms for our case are the following. When the high temperature harmonic limit has no onsite potential, then the heat conductance becomes

$$\kappa_H = \tilde{\kappa}_H \left[1 + 2\gamma_H^2 - \sqrt{1 + 4\gamma_H^2}\right]. \quad (16)$$

For the low temperature limit that has a much greater onsite term than the nearest neighbor coupling, i.e., $K_L/D_L \ll 1$, the heat conductance becomes

$$\kappa_L \approx \frac{\tilde{\kappa}_L \gamma_L^2 K_L}{D_L}, \quad (17)$$

which also assumes that the dimensionless coupling to the reservoirs is $\gamma_L \geq 1$. For strong coupling to the reservoirs, the ratio becomes

$$R \approx \frac{2K_H D_L}{K_L^2}. \quad (18)$$

This is the analytic expression that demonstrates that softening of a harmonic lattice increases the thermal conductance ratio ($D_L \propto \omega_L^2$). The strong coupling limit gives the extreme value of $R$.

Now we show that the characteristic frequency in the PBD model is lowered as $T$ crosses $T_c$ from below. For the low temperature Hamiltonian, the corresponding equation of motion is

$$m\ddot{y}_n = -\{2Da^2 y_n + K(1+\rho)[(y_n - y_{n-1}) + (y_n - y_{n+1})]\}. \quad (19)$$

From the ansatz $y_n = y_n^0 e^{i\omega t - ikn}$, one obtains the phonon spectrum as

$$m\omega^2 = 2Da^2 + 2K(1+\rho)[1 - \cos(k)]. \quad (20)$$

Thus, the frequency band of phonons is $\sqrt{2Da^2/m} \leq \omega \leq \sqrt{[2Da^2 + 4K(1+\rho)]/m}$. For the high temperature Hamiltonian, the equation of motion is

$$m\ddot{y}_n = -K[(y_n - y_{n-1}) + (y_n - y_{n+1})] \quad (21)$$

and the phonon spectrum is

$$m\omega^2 = 2K[1 - \cos(k)]. \quad (22)$$

The frequency band of phonons is $0 \leq \omega \leq \sqrt{4K/m}$. The two limiting Hamiltonians are both harmonic and the characteristic frequency is indeed lowered, and similar considerations apply for other models. In the low temperature limit, the onsite potential stiffens the DNA compared to the high temperature limit, which results in the raising of the phonon spectrum of the low temperature limit compared to the high temperature. In the latter, as well, the nearest neighbor coupling drops from $K(1+\rho)$ to $K$ shrinking the bandwidth. This trade-off is responsible for the change in thermal conductance across the transition. If the drop in nearest neighbor coupling is small, then the softening will dominate, and the heat conductance will increase because these softened modes can conduct heat more effectively.

Next we consider a lattice with periodic segments. For a periodic lattice with alternating (AG) pairs, $N_s = 2$ so there are two onsite parameters $D_\mu^{1,2}$ and $\mathcal{T}_s = \mathcal{T}_1 \mathcal{T}_2 = \begin{pmatrix} \mathcal{D}_{1,2}^0 & -\mathcal{D}_{2,2}^0 \\ \mathcal{D}_{1,1}^0 & -\mathcal{D}_{2,1}^0 \end{pmatrix} = \begin{pmatrix} [x_1] & -1 \\ 1 & 0 \end{pmatrix} \begin{pmatrix} [x_2] & -1 \\ 1 & 0 \end{pmatrix}$, where $[x_1] \equiv 2(1 + \frac{D_\mu^1}{K_\mu}) - \frac{m}{K_\mu}\omega^2$ and $[x_2] \equiv 2(1 + \frac{D_\mu^2}{K_\mu}) - \frac{m}{K_\mu}\omega^2$. From Eq. (8) and $|\cos(q)| \leq 1$ one can find two bands which satisfy the condition:

$$u_{b1} \leq \omega^2 \leq u_{b2} \text{ and } u_{t1} \leq \omega^2 \leq u_{t2}. \quad (23)$$

Here $u_{b1} = (K_\mu/m)[2 + (D_\mu^1 + D_\mu^2)/K_\mu - \sqrt{4 + (D_{\mu 1} - D_\mu^2)^2/K_\mu^2}]$, $u_{b2} = 2(K_\mu/m)(1 + D_\mu^1/K_\mu)$, $u_{t1} = 2(K_\mu/m)(1 + D_\mu^2/K_\mu)$, and $u_{t2} = (K_\mu/m)[2 + (D_\mu^1 + D_\mu^2)/K_\mu + \sqrt{4 + (D_\mu^1 - D_\mu^2)^2/K_\mu^2}]$.

One can show that the current is

$$\frac{J}{\Delta T} = \frac{\lambda}{\pi} \frac{m^2}{K_\mu^2} \int_{\omega \in W} d\omega |\omega \sin(q)| [(1 + \lambda^2 \omega^2 m^2/K_\mu^2) \times$$
$$(m/K_\mu)|\mathcal{D}_{1,1}^0 + \mathcal{D}_{2,2}^0|]^{-1}$$
$$= \frac{\lambda m}{2\pi K_\mu} \left( \int_{u_{b1}}^{u_{b2}} + \int_{u_{t1}}^{u_{t2}} \right) du \sqrt{1 - \cos^2(q)} [(1 +$$
$$\lambda^2 m^2 u/K_\mu^2)(4 + 2D_\mu^1/K_\mu + 2D_\mu^2/K_\mu -$$
$$2mu/K_\mu)]^{-1}. \quad (24)$$

Here we have used $u = \omega^2$ and $\cos(q(u)) = (1/2)[(2(1 + D_\mu^1/K_\mu) - mu/K_\mu)][2(1 + D_\mu^2/K_\mu) - mu/K_\mu] - 1$ from Eq. (8).

The heat currents for more complicated bases can be derived in a similar way. For example, the heat current for a poly($A_2G$) strand is given by

$$\frac{J}{\Delta T} = \frac{\lambda m}{2\pi K_\mu} \int_{\omega \in W} du \sqrt{1 - \cos^2(q)} [(1 +$$
$$\lambda^2 m^2 u/K_\mu^2)|[x_1][x_2] + [x_1]^2 - 2|]^{-1} \quad (25)$$

Here the three conduction bands are determined by $2\cos(q) = \mathcal{D}_{1,3}^0 - \mathcal{D}_{2,2}^0 = [x_1]^2[x_2] - 2[x_1] - [x_2]$. The heat current for a poly($A_2G_2$) strand is

$$\frac{J}{\Delta T} = \frac{\lambda m}{2\pi K_\mu} \int_{\omega \in W} du \sqrt{1 - \cos^2(q)} [(1 + \lambda^2 m^2 u/K_\mu^2) \times$$
$$|[x_1]^2[x_2] + [x_1][x_2]^2 - 2[x_1] - 2[x_2]|]^{-1} \quad (26)$$

The four conduction bands are determined by $2\cos(q) = \mathcal{D}_{1,4}^0 - \mathcal{D}_{2,3}^0$, where $\mathcal{D}_{1,4}^0 = [x_1]^2[x_2]^2 - [x_1][x_2] - [x_1]^2 - [x_2]^2 + 1$, $\mathcal{D}_{2,3}^0 = [x_1][x_2] - 1$, $\mathcal{D}_{1,3}^0 = [x_1][x_2]^2 - [x_1] - [x_2]$, and $\mathcal{D}_{2,4}^0 = [x_1]^2[x_2] - [x_1] - [x_2]$.

For parameters relevant to real DNA, we have evaluated the thermal conductance (in the harmonic limit) of poly(A), poly(G) homogeneous lattices as well as poly(AG), poly($A_2G$), poly($AG_2$) and poly($A_2G_2$). They qualitatively agree with our numerical simulations.

### III. SIMULATION DETAILS

#### A. Transfer matrix

The transfer matrix formalism is based on the possibility to evaluate the partition function for a classical non-linear lattice using the matrix algebra [9]. Specifically, the total classical partition function for the PBD model is given by

$$\mathcal{Z} = \mathcal{Z}_T \mathcal{Z}_U, \quad (27)$$

where $\mathcal{Z}_T = (2\pi m k_B T)^{N/2}$ is the "kinetic" partition function and the "potential" partition function reads as

$$\mathcal{Z}_U = \int \prod_{n=1}^N dy_n \, e^{-[V_n(y_n) + W_n(y_n, y_{n-1})]/k_B T}. \quad (28)$$

The continuous integration can be represented through summation on the grid yielding for periodic boundary conditions (i.e., $y_0 \equiv y_N$)

$$\mathcal{Z}_U = \text{Tr}\left[\prod_{n=1}^{N} \hat{M}_n\right], \quad (29)$$

where matrix $\hat{M}_n$ is defined through its matrix elements as

$$[\hat{M}_n]_{ij} = \Delta x e^{\left[\frac{1}{2}V_n(x_i) + W_n(x_i,x_j) + \frac{1}{2}V_{n-1}(x_j)\right]/k_B T}, \quad (30)$$

where $x_i$ is the coordinate on the grid with the grid step $\Delta x$, i.e., $x_i = x_0 + i\Delta x$. The parameters of the grid used in our calculations are $x_{min} = -1.5$ Å, $x_{max} = 50$ Å with $\Delta x = 0.05$ Å.

In the case of a homogeneous lattice, i.e., if $V_n(y_n) \equiv V(y_n)$ and $W_n(y_n, y_{n-1}) \equiv W(y_n, y_{n-1})$, all the matrices $\hat{M}_n$ are identical and symmetric, which results in the possibility to evaluate the trace in Eq. (29) as $\mathcal{Z}_U = \sum_i \lambda_i^N$, where $\lambda_i$ are the eigenvalues of symmetric matrix $\hat{M}_n$. Naturally, if the lattice becomes very long, only the largest eigenvalue, $\lambda_{max}$ of the transfer matrix contributes to the partition function yielding $\mathcal{Z}_U = \lambda_{max}^N$

The natural generalization of this procedure to the case of DNA strand made of repeating motif is as follows. First, matrices corresponding to a single repeating unit $(r)$ are multiplied yielding $\hat{M}_r = \prod_{n \in r} \hat{M}_n$. The potential partition function is given by $\text{Tr}[\hat{M}_r^{N_r}]$, where $N_r$ is the number of repeat units in the strand with periodic boundary conditions. Therefore, the potential partition function becomes $\mathcal{Z}_U = \sum_i v_i^{N_r}$, where $v_i$ are the eigenvalues of $\hat{M}_r$. We note here, that even though matrix $\hat{M}_r$ might not be symmetric, the eigenvalue problem for such a matrix is still well defined and eigenvalues is all what is needed to evaluate the "potential" partition function above. Specifically, it can be shown that the Schur decomposition, which yields eigenvalues, is always possible for any square matrix. This decomposition does not guarantee the completeness of the eigenvector-based basis, but *only* knowledge of eigenvalues is required for the evaluation of the potential partition function for the lattice with periodic boundary conditions.

Once the partition functions are found, the heat capacity of the DNA strand, normalized per number of sites, can be evaluated as

$$C = -\frac{T}{N}\frac{\partial^2 F}{\partial T^2}, \quad (31)$$

where $F = -k_B T \ln(\mathcal{Z}_T \mathcal{Z}_U)$ is the free energy of the DNA strand.

### B. Langevin dynamics

To study the dynamics of the DNA out of equilibrium we solve numerically the Langevin equation, which de-

|    | $K(\text{eV}/\text{Å}^2)$ |
|----|-----------|
| $AA$ | 0.023 |
| $GG$ | 0.019 |
| $AG$ | 0.0232 |
| $GA$ | 0.0185 |

Table I. Harmonic parameter of the sequence-dependent stacking interaction of the PBD model.

scribes the dynamics of a Hamiltonian system in the presence of thermal baths. The Langevin equation is given by

$$m\ddot{y}_n = -\frac{\partial W}{\partial y_n} - \frac{\partial V}{\partial y_n} - \Gamma_n \dot{y}_n + f(t), \quad (32)$$

where $W(y_n)$ and $V(y_n)$ are the potentials described in Eq. (1) of the main text. The DNA strand is split into three regions, the two ends, each of length $l$, serve as the Langevin thermal reservoirs at temperatures $T_L$ and $T_H$. This means that the friction term $\Gamma_n$ only operates for $n$ within the thermal reservoirs. The fluctuating term $f(t)$ is Gaussian white noise which obeys the fluctuation-dissipation relation $\langle f(t)f(t')\rangle = 2\Gamma_n k_B T_{L(H)} \delta(t-t')$ for the low and high temperature reservoirs, respectively. In our simulations, the parameters of each Langevin thermostat have always been set to $l = 20$ and $\Gamma_n/m = 0.5$ ps$^{-1}$.

The middle region of the length $M$ is the free DNA strand, which is driven out of equilibrium by the Langevin reservoirs when $T_L \neq T_H$. The length of the middle region has been varied in the range $M = 50 - 200$ in our simulations. The parameters of the PBD model are given in the main text except for the sequence-dependent stacking interaction strength. These are adopted from Ref. 5 and compiled in Table I. The equations of motion are integrated with the fourth-order Runge-Kutta method. The local heat current is given by $J_n = -\left\langle \dot{y}_n \frac{\partial W(y_n, y_{n-1})}{\partial y_n}\right\rangle$. The simulations are performed long enough to allow the system to reach its steady state, where $J_n$ does not depend on $n$, as long as $n$ is within the "free" middle region of DNA.



\end{document}